\documentclass[
reprint,
superscriptaddress,
 amsmath,amssymb,
 aps,
prb,
floatfix,
]{revtex4-2}
\usepackage{multirow}
\usepackage{graphicx}
\usepackage{dcolumn}
\usepackage{bm}
\usepackage[english]{babel}
\usepackage{hyperref}

\begin{document}

\title{A Kitaev--type spin liquid on a quasicrystal}

\author{M. A. Keskiner}
\email{akif.keskiner@bilkent.edu.tr}
\affiliation{Department of Physics, Bilkent University, Ankara, 06800, TURKEY}
\author{O. Erten}
\email{onur.erten@asu.edu}
\affiliation{Department of Physics, Arizona State University, Tempe, AZ 85287, USA}
\author{M. \"{O}. Oktel}
\email{oktel@fen.bilkent.edu.tr}
\affiliation{Department of Physics, Bilkent University, Ankara, 06800, TURKEY}

\date{\today}

\begin{abstract}
We develop an exactly solvable model with Kitaev-type interactions and study its phase diagram on the dual lattice of the quasicrystalline Ammann-Beenker lattice. Our construction is based on the $\Gamma$-matrix generalization of the Kitaev model and utilizes the cut-and-project correspondence between the four-dimensional simple cubic lattice and the Ammann-Beenker lattice to designate four types of bonds. We obtain a rich phase diagram with gapped (chiral and abelian) and gapless spin liquid phases via Monte Carlo simulations and variational analysis. We show that the ground state can be further tuned by the inclusion of an onsite term that selects 21 different vison configurations while maintaining the integrability of the model. Our results highlight the rich physics at the intersection of quasicrystals and quantum magnetism.

\end{abstract}

\maketitle

\section{Introduction}
Quantum spin liquids (QSLs) are the disordered phases of magnetic systems and exhibit exceptional properties such as fractionalized excitations and long-range entanglement due to their underlying topological order\cite{Zhou,Clark,Wen2019,Wen, Balents_Nature2010, Wen_RMP2017, Knolle_AnnRevCondMatPhys2019, Broholm_Science2020}. The Kitaev model on the honeycomb lattice is a foundational model as it is the first exactly solvable spin model that showcases a spin liquid ground state with both gapless and gapped phases, featuring abelian and non-abelian anyonic excitations\cite{KITAEV20062}. Recently, there has been a tour de force effort in discovering new materials with strong Kitaev interactions such as iridates and $\alpha-$RuCl$_3$\cite{HwanChun_NatPhys2015, Takagi_NatRevPhys2019,Kitagawa_Nat2018}. Kitaev interactions may be strong in other van der Waals (vdW) materials like CrI$_3$ \cite{Lee_PRL2020}.

Most models of quantum magnetism, particularly models exhibiting a QSL ground state, have been explored in the context of periodic systems.  While QSLs lack long-range magnetic order, they are in general defined on models with perfect translational symmetry. 
For instance, recent works have explored the consequences of translational symmetry breaking for the QSL state, either by considering defects in a periodic lattice\cite{Kao_AnnPhys20221, Dandas_PRL2022, Kao_PRB2022, Singhania_PRR2023} or by defining randomly generated amorphous lattices\cite{Cassella,Grushin}. Solids do not have to be either periodic structures or random glasses. A third possibility is quasiperiodicity.

Quasicrystals are unique classes of materials that exhibit a regular atomic structure with non-repeating patterns, forming a contrast between the disordered arrangement of glasses and the fully periodic arrangement of crystals. They can be characterized by the presence of long-range order in terms of their translational and orientational symmetries, but the patterns do not repeat at regular intervals\cite{she84,LevineA,Levine}. Due to this unique arrangement of the sites, they exhibit rare features such as strictly-localized states \cite{Arai,Oktel4}, or electronic states which are neither localized nor extended \cite{sut86,kka14}. Quasicrystals can also exhibit different symmetries, including five- and eight-fold rotations, forbidden in conventional crystal structures. These forbidden symmetries lead to unprecedented topological states that are not allowed in crystalline materials\cite{Varjas,Fan,Chen}. 
 \begin{figure}[!t]
    \includegraphics[width=0.5\textwidth]{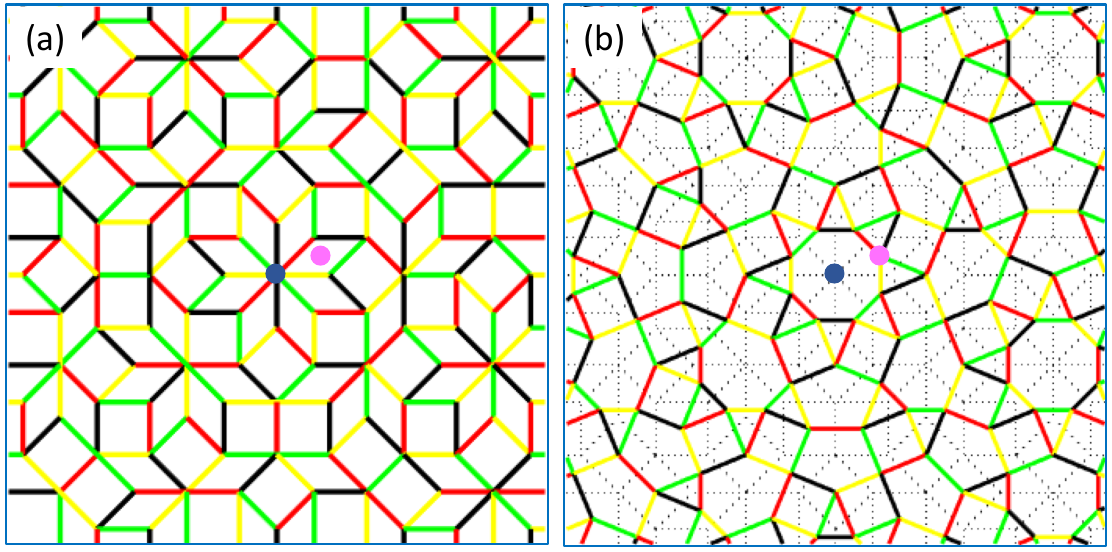}
    \caption{(a) Ammann-Beenker lattice (ABL) and the coloring scheme consisting of four type of bonds obtained via projecting from the hyperspace. (b) The dual lattice of ABL (dABL) and the corresponding coloring scheme. The pink and blue dots show the position of the lattice sites in the ABL and dABL respectively.}
    \label{fig:ColorDualLat}
\end{figure}

The understanding of the elementary excitation spectrum of quasicrystals is quite limited compared to periodic solids. Experimentally, most quasicrystals are found to be poor conductors \cite{poo92}. However, non-Fermi liquid behavior\cite{dms12} and superconductivity \cite{kam18} have also been observed. 
Recent experiments with high-quality samples have demonstrated that ferromagnetic long-range order is possible in a quasicrystal \cite{tis21}. The experimental realization of quasicrystalline systems goes beyond the synthesis of alloys. Vapor deposition on metallic surfaces\cite{fmh13}, as well as atomic force microscopy assembly of molecules \cite{col17} have created large two-dimensional quasicrystals. Synthetic quasicrystals for light \cite{var13}, and ultracold atoms \cite{vie19} promise to probe bosonic models in quasicrystals. In a more recent development, quasicrystals can also be formed in moir\'e superlattices of vdW materials\cite{Uri_arXiv2023}.
Experimental possibilities for testing the excitation spectrum of quasicrystals are rapidly increasing. 

Classical models of magnetism, such as the Ising model, have been studied in some detail for quasicrystals \cite{koo16,thc15}. The non-trivial structural properties of quasicrystals force a reconsideration of the basic models, such as dimer coverings on the simplest two-dimensional lattices \cite{fsp20,lbs22}. However,  quantum magnetism research in quasicrystals is at its early stages. Numerical work on small systems \cite{kon05} and two-dimensional lattices \cite{wjh03,jsw07,kts17} indicate elements of frustration and self-similar magnetic ground states. Recently,
Ref.~\citenum{Ghosh} showed that exact dimer wave functions can be constructed in certain quasicrystals for generalized Heisenberg Hamiltonians. 

Here, we investigate the interplay between quasiperiodicity and spin liquid order by constructing an exactly solvable model with Kitaev-type interactions on a quasicrystal. The Kitaev model and its generalizations are integrable due to the presence of conserved quantities at each mesh of the lattice. The conservation of such mesh loops requires two structural properties. First, the lattice must have the same coordination number $z$ on each site. Second, the link lattice itself must be $z-$partite, i.e., the links of the lattice can be labeled by one of the $z$ colors so that all links starting from any site have different colors.  The two most common quasicrystal models, the Penrose \cite{pen74} (PL) and Ammann-Beenker \cite{amm92,bee82} (ABL) lattices, have varying coordination numbers for their vertices. However, both of these models have quadrilateral tiles and their dual lattices have a constant coordination number $z=4$. The tight-binding models on the dual lattices are also referred to as the center tight-binding models \cite{fat88}. Although the condition of uniform coordination is easily satisfied by considering the dual lattices, the second condition for the z-partite coloring of the lattice is non-trivial.  For the dual of the Ammann-Beenker lattice (dABL), we find a z-partite coloring rule by lifting the ABL back into the four-dimensional space. We show that the four-dimensional cubic lattice with nearest neighbor links can be colored in a way so that each square has sides of 4 different colors. This rule allows us to define the Kitaev-type Hamiltonian on the dABL. It is worth remarking that the same method cannot be extended to a five-dimensional simple cubic lattice, and we have not been able to define a similar model on the PL.

\begin{figure*}[t]
    \centering
    \includegraphics[clip,width=1\textwidth]{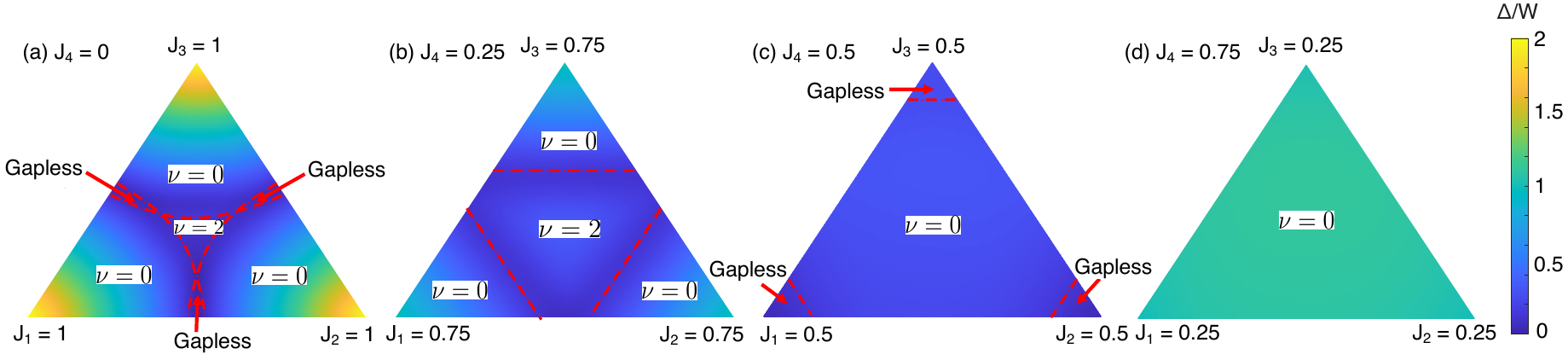}
    \caption{Ground state phase diagram for $J_1+J_2+J_3+J_4=1$ and $J_5=0$. (a), (b), (c), (d) panels have $J_4$=0, 0.25, 0.5, and 0.75 respectively. The gap value is normalized by the bandwidth which varies depending on the parameters. The phase diagram contains two gapped phases with $\nu=2$ and $\nu=0$ which are separated by a gapless region.}
    \label{fig:Phasediagrams}
\end{figure*}

Our coloring scheme for the ABL obtained by projection and corresponding dABL are shown in Fig. \ref{fig:ColorDualLat}(a), and Fig. \ref{fig:ColorDualLat}(b). In this model, an exact solution of the spin model can be achieved via a Majorana fermion representation of the $\Gamma$ matrices. dABL consists of plaquettes with both even- and odd-numbered sides. The odd-numbered plaquettes break time reversal symmetry\cite{KITAEV20062} and allow for chiral spin liquid phases that are classified by a Chern number, $\nu$. We obtain the ground state phase diagram via Monte Carlo simulations and variational analysis. We find that the ground state follows a flux configuration that is an extension of the Lieb's theorem \cite{Lieb}, even though the theorem was not proved for non-bipartite lattices. Depending on the coupling constants ($J_\mu,~\mu =1,2,3,4$), we find that the ground state can be gapped with $\nu=2$ or $\nu=0$. These phases are separated by a gapless phase, as shown in Fig.~\ref{fig:Phasediagrams}. Upon inclusion of an onsite term that commutes with the flux operators, the ground state vison configuration changes as shown in Fig. \ref{fig:Phase_diag_j1111}. We find that up to 21 different flux configurations can be stabilized as a function of the onsite field strength.

The paper is organized as follows. Section II briefly overviews the Ammann-Beenker lattice and its periodic approximants.  Section III focuses on constructing our model with Kitaev-type interactions, including our coloring scheme.  In section IV, we present our results and discuss the phase diagram. We conclude with a discussion and a summary of our results.

\section{Ammann-Beenker Lattice}

We use one of the most well-known quasicrystals, the Ammann-Beenker Lattice (ABL), to construct our model. All the points forming the ABL can be written as
\begin{equation}
\label{eq:lattice}
\vec{R}_{ABL}= k_0 \hat{e}_0 +k_1 \hat{e}_1+k_2 \hat{e}_2+k_3 \hat{e}_3,
\end{equation}
where  $k_j$ are integers and the star-vectors are
\begin{equation}
    \hat{e}_0 = \hat{x}, \,
    \hat{e}_1 = \frac{1}{\sqrt{2}}\left(\hat{x}+\hat{y}\right), \,
    \hat{e}_2 = \hat{y},\,
    \hat{e}_3 = \frac{1}{\sqrt{2}}\left(-\hat{x}+\hat{y}\right).
\end{equation}
All the bonds in the ABL are parallel to one of the above star vectors, making the meshes of ABL either squares or $\pi/4$ rhombi.  

If the integers $k_j$ in Eq.~\ref{eq:lattice} are allowed to vary independently, the set of lattice points will uniformly fill the plane and create points that are infinitesimally close to each other. The definition of the quasicrystal lattice can be seen as a way of constraining the set $k_0,k_1,k_2,k_3$. Although alternative methods exist for defining this constraint, we use the cut-and-project method which relates the ABL to the four-dimensional cubic lattice. Each quadruple $k_0,k_1,k_2,k_3$, defines a unit tessaract. ABL is formed by using $k_j$ which has an intersection with a specific two-plane. The two-plane is chosen so that the projection of unit vectors in four dimensions onto this plane defines the star vectors given above \cite{bee82}. 

The first condition necessary for an exactly solvable Kitaev model is to obtain a lattice with a uniform coordination number. ABL has vertices with coordination numbers varying from 3 to 8, hence does not satisfy this condition. However, as each mesh of ABL is a quadrilateral, its dual lattice (dABL) is uniformly coordinated. dABL is obtained by placing vertices to the center of each mesh and forming a link between two sites if their meshes share an edge. So each dABL site is expressed as $\vec{R}_{dABL}=\vec{R}_{ABL}\pm \frac{\hat{e}_m}{2}\pm \frac{\hat{e}_n}{2}$, where the edges of the mesh including this vertex are parallel to $\hat{e}_{m}$ and $\hat{e}_{n}$.  
Tight binding models have been studied on dABL and other dual lattices, where such models are referred to as center models \cite{fat88}.

The second condition we need to define a Kitaev-type model is to make sure that the uniformly coordinated dABL is 4-partite. Each of the links in the dABL has to be assigned an index from 1 to 4, with the condition that all 4 links meeting at every lattice site has a different index. The same condition can be expressed in terms of the direct ABL as coloring all the links so that each mesh has 4 edges of different colors. By considering the four-dimensional simple cubic lattice, we establish a coloring rule that satisfies this condition. As the cut-and-project construction shows every mesh in the ABL corresponds to a mesh in a two-plane of the four dimensional cubic lattice. The details of this coloring scheme for the four-dimensional lattice which satisfies the coloring condition is presented in the Appendix~\ref{sec:Coloring}. The resulting coloring for the ABL and dABL is shown in Fig.~\ref{fig:ColorDualLat}. Our coloring rule cannot be trivially generalized to higher dimensions, hence to other quasicrystals such as the PL. 

\begin{figure}[t]
    \includegraphics[clip,width=0.5\textwidth]{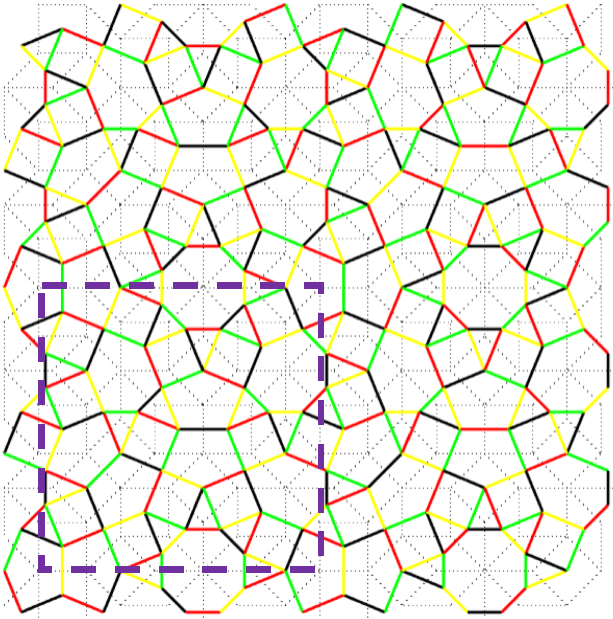}
    \caption{Four unit cells of the approximant, represented by s=2 for dABL. The black dotted lines show the original ABL and the purple dashed lines show a single unit cell of the 
    s=2 approximant.}
    \label{fig:Approximant}
\end{figure}

While our model is defined for the infinite quasicrystal, numerical calculations require a finite region. We use two kinds of regions for our calculations. First, we use a finite region with open boundary conditions. We choose different local configurations for the finite region to make sure that our results are not specific to a particular patch of the quasicrystal. We use a periodic approximant for the ABL \cite{Duneau} as the second method. Periodic approximants are obtained by choosing a cut plane with a rational slope in the cut-and-project construction. The approximants are indexed by an integer $s$ which increases the number of sites within the unit cell and creates the infinite ABL in the limit $s\rightarrow \infty$. For $s=1,2,3$  the periodic unit cell has $N=7,41,239$ sites respectively.  The unit cell for the ABL approximants is a square, and we double the unit cell in both directions for compatibility with our coloring rules. In  Fig.~\ref{fig:Approximant} we display the enlarged unit cell for the $s=2$ approximant with the dABL coloring.

\begin{figure}[!t]
    \centering    \includegraphics[clip,width=0.5\textwidth]{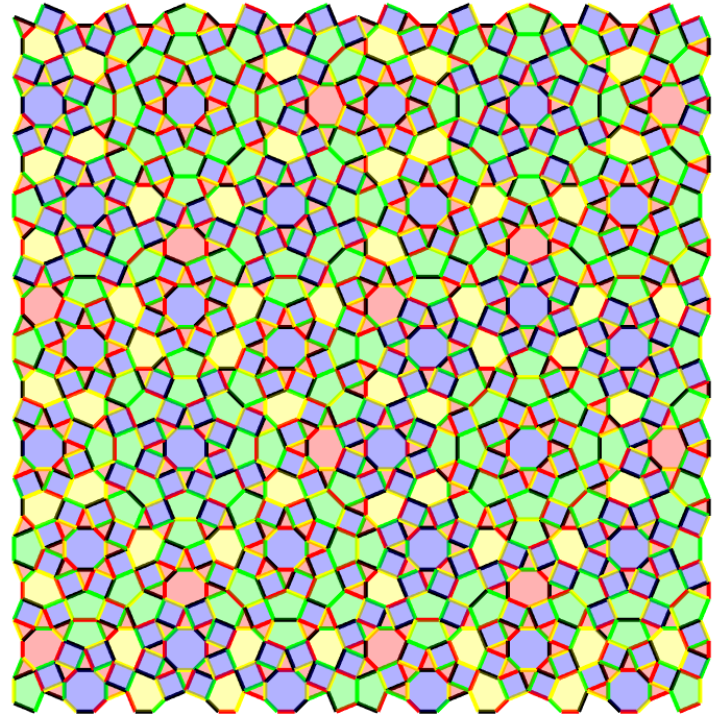}  
    \caption{Ground state flux configuration for $J_5=0$ for a s=3 approximant. The $0$-flux, $\pi$-flux, $\frac{\pi}{2}$-flux, and $-\frac{\pi}{2}$-flux are represented by yellow, blue, red, and green colors, respectively. The ground state breaks time-reversal symmetry due to odd-numbered plaquettes.}
    \label{fig:Lieb}
\end{figure}
\section{Microscopic Model}
An essential requirement for solving the Kitaev model exactly is the anticommutation relations of the Pauli matrices, $\{\sigma_{i}, \sigma_j \} = 2\delta_{ij}$. Since there are only three Pauli matrices, this approach is applicable only to lattices with a coordination number of $z = 3$, such as the honeycomb, hyperhoneycomb, and hyperoctagon lattices. However, an extension of Kitaev's method is possible using $\Gamma$ matrices that follow the Clifford algebra $\{\Gamma_{i}, \Gamma_j \} = 2\delta_{ij}$ \cite{Gamma,Yao}. For instance, in a four-dimensional representation of the Clifford algebra, there exist five $\Gamma^\mu$ operators, along with ten $\Gamma^{\mu \nu}=\frac{i}{2}[\Gamma^{\mu},\Gamma^{\nu}]$ operators and an identity matrix, which form the basis of the local Hilbert space. Consequently, Kitaev's construction can be expanded to lattices with a coordination number of up to $z = 5$ \cite{Gamma,Yao}. dABL has coordination number $z=4$, represented by four different colors for each type of bond. Therefore, we use this representation and consider the Hamiltonian,
\begin{eqnarray}
H_K=-\sum_{\langle ij\rangle_{\mu}} J_\mu (\Gamma_{i}^{\mu}\Gamma_{ j}^{\mu}+\Gamma_{ i}^{\mu5}\Gamma_{ j}^{\mu5})+J_{5}\sum_{j}\Gamma_{j}^{5}
\end{eqnarray}
where $\mu$ is the type if the bond, $\mu = \{1,2,3,4\}$. Since the $\Gamma^5$ operator is not used as a bond operator, it can be included as an onsite term. Similar models have been considered for other $z=4$ lattices such as square lattice\cite{NRF, Urban, Yao}. Note that four-dimensional $\Gamma$ matrices can be represented by two sets of Pauli matrices \cite{Yao_PRL2011, Erten, Nica_npjQM2023} with Kugel-Khomskii interactions\cite{Kugel_SovPhys1982} or $J=3/2$ operators\cite{Yao}. For the latter, $\Gamma^\mu$ operators correspond to quadrupolar operators. For each plaquette, we define  $W_p=\prod_{(j,k)\in p}\Gamma_j^{\mu} \Gamma_k^{\mu} $ where the product is taken in the counter-clockwise direction. These plaquette operators commute with the Hamiltonian, $[H, W_p]=0$ and each other $[W_p, W_{p^\prime}]=0$. Therefore this leads to infinitely many conserved quantities for the model. An exact solution can be obtained by a Majorana representation of the $\Gamma$ matrices \cite{Gamma},
\begin{equation}
    \Gamma_j^{\mu}=ib_j^{\mu}c_j,~~~~\Gamma_j^{\mu\nu}=ib_j^{\mu}b_j^{\nu}
\end{equation}
where we introduce a total of 6 Majorana fermions per site. Relabeling $b_{j}^{5}$ as $b_{j}^{5}\rightarrow d_j$, the Hamiltonian can be reexpressed as
\begin{eqnarray}
H= \sum_{\langle ij\rangle_{\mu}}J_\mu iu_{ij}^{\mu}(c_ic_j+d_{i}d_{j})+
    J_{5}\sum_{i}id_{i}c_i
    \label{eq:Hmf}
\end{eqnarray}
where $u_{ij}^\mu=ib_{i}^{\mu}b_{j}^{\mu}$. $u_{ij}^\mu$ also commute with the Hamiltonian, $[H, u_{ij}^\mu] = 0$ and thus it is also a constant of motion. This representation is redundant, and the physical states need to be eigenstates of $D_{ i}=ib^1_{i}b^2_{i}b^3_{i}b^4_{i}c_{i}d_{i}$ with eigenvalues $+1$. These constraints can be implemented by a projection operator $P=\prod_i(1+D_{i})/2$. A $\mathbb{Z}_{2}$ gauge transformation at site $i$ involves flipping the signs of the Majorana fermions and bond operators, $\{c_{i}, d_i \} \rightarrow \{-c_{i}, -d_i \};~ u^\mu_{ij} \rightarrow -u^\mu_{ij}$. The plaquette operators can be expressed in terms of the bond operators as $W_p=(-i)^n\prod_{(j,k)\in p}u_{jk}^\mu$ where n is the number of the links on the boundary of the plaquette p and the product is taken in the counter-clockwise direction. The eigenvalues of $W_p$ for even (odd) plaquettes can take values $\pm 1$ ($\pm i$). Therefore the solution of the model involves two flavors of free Majorana fermions hopping in the background of static $\mathbb{Z}_2$ fluxes. Since there is no Lieb's theorem for quasicrystals, we obtain the ground state phase diagram via Monte Carlo simulations for approximants on varying sizes, s=1, 2, and 3. We use the Metropolis algorithm for the flux degrees of freedom, and for a given vison configuration, we perform exact diagonalization to obtain the corresponding energy. We perform 5000 Monte Carlo steps for a given temperature and reduce the temperature down to $10^{-3}J$. Motivated by the Monte Carlo results, we also construct various variational states and compare their energies. For instance, in the case of finite $J_5$ calculations, we construct 32 vison configurations for which 21 of them are stabilized as a function of $J_5/J$.

\section{Results and Discussion}
We first discuss the phase diagram in the absence of the onsite anisotropy term ($J_5=0$). Even though Lieb's theorem\cite{Lieb} does not apply to quasicrystals, we observe that the ground state follows a simple rule that is an extension of Lieb's theorem, and the fluxes through the plaquettes, $W_p$, are given by
\begin{eqnarray}
    \phi_p = - (\pm i)^n
    \label{eq:flux}
\end{eqnarray}
which recovers the well-known result that honeycomb (square, octagon) plaquettes exhibit 0 ($\pi$) fluxes whereas the triangle, heptagon (pentagon) plaquettes have $\pm \pi/2$ ($\mp \pi/2$) fluxes in the ground state as shown in Fig.~\ref{fig:Lieb}. We will refer this vison configuration `L' for the remainder of the article. Albeit there is no proof for the extension of Lieb's theorem for graphs that has plaquettes with varying number of edge sites, n, similar behaviors have also been observed in amorphous and polycrystalline Kitaev models\cite{Cassella,Grushin} and Kitaev models in closed-geometries (i. e. tetrahedrons)\cite{Mellado_PRB2015}. Eq.~\ref{eq:flux} implies that the ground state flux configuration spontaneously breaks time-reversal symmetry due to the odd-numbered plaquettes (n=3, 5, 7). This phenomenon was originally pointed out by Kitaev\cite{KITAEV20062} and was first discovered in a model by Yao and Kivelson\cite{Yao}. A time reversal operation on a ground state flips the sign of the fluxes on the odd-numbered plaquettes and generates a new ground state. 
\begin{figure}[!t]
    \centering
    \includegraphics[clip,width=0.4\textwidth]{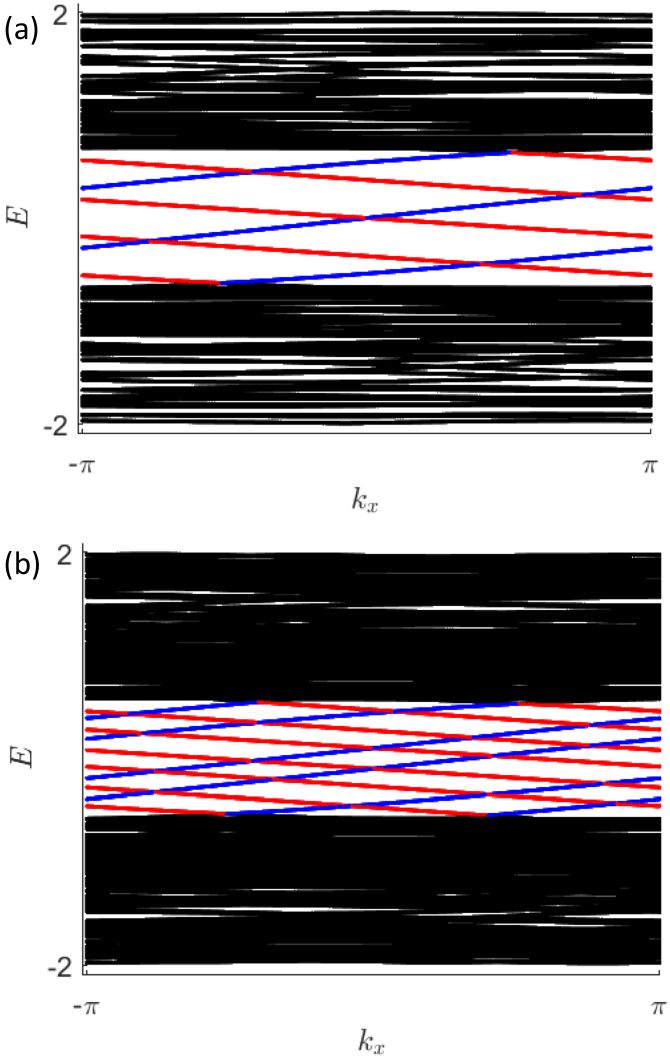}
    \caption{(a) Edge states without the onsite term for a point in the phase space characterized by the Chern number 2. (b) Edges states for the same point with onsite term splits the edge states but preserves $\nu=2$.}
    \label{fig:Edges}
\end{figure}
\begin{figure*}[!t]
    \centering
    \includegraphics[clip,width=0.9\textwidth]{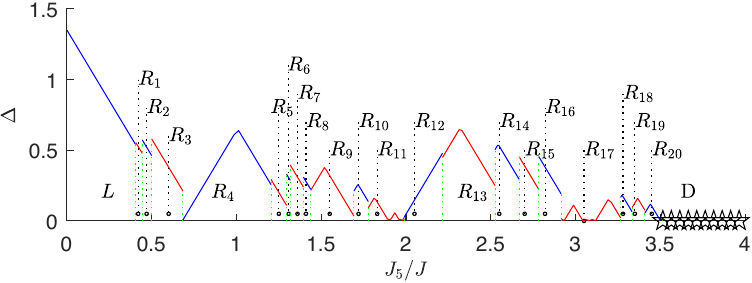}
    \caption{Phase diagram as a function of the onsite anisotropy, $J_5/J$ for $J_1=J_2=J_3=J_4=J$. The y-axis represents the gap, $\Delta$. We obtain 21 different vison configurations shown in Table I.}
    \label{fig:Phase_diag_j1111}
\end{figure*}

We find that the L vison configuration remains the ground state even with anisotropic couplings, $J_\mu$, as long as $J_5=0$. However, similar to Kitaev's original model, the Majorana band structure depends on the values of $J_\mu$'s. To investigate that, we study the phase diagram for $J_1+J_2+J_3+J_4=1$, which forms a tetrahedron phase space. For the sake of a simpler and clearer presentation, we take four cuts on the tetrahedron as shown in Fig.~\ref{fig:Phasediagrams}. We find two gapped phases: a chiral spin liquid with $\nu=2$ and an abelian spin liquid with $\nu=0$. The chiral edge modes for the $\nu=2$ phase originate from the two flavors of Majorana fermions both contributing a single chiral edge mode. For $J_5=0$, the two Majorana fermions are completely decoupled and have an identical spectrum. We identify the Chern number by counting the chiral edge modes on a slab geometry. An example of the chiral edge modes are shown in Fig.~\ref{fig:Edges}. The gapped phases are connected by a gap-closing transition and in certain cases a finite region of the gapless phase (see Fig~\ref{fig:Phasediagrams}(a). We note that our phase diagram shows similarities to the amorphous Kitaev model\cite{Cassella,Grushin} which has $\nu=0$ and $\nu=1$ phases. 

\begin{table}[t!]
\begin{tabular}{|c|c |c |c |c |c |c|} 
 \hline
 Flux sector& $n=3$ & $n=4$& $n=5$&$n=6$ &$n=7$&$n=8$\\ [0.5ex] 
 \hline\hline
 $L$&$\pi/2 $ & $\pi$ & $-\pi/2 $ & 0 &$+\pi/2 $ & $\pi$  \\ 
 \hline
  $R_1$&$\pi/2  $& $\pi$ & $-\pi/2 $ & 0 &$+\pi/2 $ & 0  \\ 
 \hline
  $R_2$&$\pi/2 $ & $\pi$ & $-\pi/2 $ & 0 &$-\pi/2 $ & 0  \\ 
 \hline
  $R_3$&$\pi/2 $& $\pi$ & $-\pi/2 $ & $\pi$ &$-\pi/2 $ & 0  \\ 
 \hline
  $R_4$&$\pi/2 $ & $\pi$ & $+\pi/2 $ & $\pi$ &$-\pi/2 $ & 0  \\ 
 \hline
  $R_5$&$\pi/2  $& $\pi$ & $+\pi/2 $ & $\pi$ &$-\pi/2 $ & $\pi$  \\ 
 \hline
  $R_6$&$\pi/2 $ & $\pi$ & $+\pi/2 $ & $\pi$ &$+\pi/2 $ & $\pi$  \\ 
 \hline
  $R_7$&$\pi/2 $ & $0$ & $-\pi/2 $ & 0 &$-\pi/2 $ & 0  \\ 
 \hline
  $R_8$&$\pi/2 $ & $0$ & $-\pi/2 $ & 0 &$-\pi/2 $ & $\pi$  \\ 
 \hline
  $R_{9}$&$\pi/2 $ & $0$ & $-\pi/2 $ & $\pi$ &$-\pi/2 $ & $\pi$  \\ 
 \hline
  $R_{10}$&$\pi/2 $ & $0$ & $+\pi/2 $ & 0 &$-\pi/2 $ & $\pi$  \\ 
 \hline
  $R_{11}$&$\pi/2  $& $0$ & $+\pi/2 $ & $\pi$ &$-\pi/2 $ & 0  \\ 
 \hline
  $R_{12}$&$\pi/2 $ & $0$ & $+\pi/2 $ & $\pi$ &$+\pi/2 $ & $\pi$  \\ 
 \hline
  $R_{13}$&$\pi/2 $ & $0$ & $+\pi/2 $ & $\pi$ &$+\pi/2 $ & 0  \\ 
 \hline
  $R_{14}$&$\pi/2 $ & $0$ & $+\pi/2 $ & 0 &$+\pi/2 $ & 0  \\ 
 \hline
  $R_{15}$&$\pi/2 $ & $0$ & $+\pi/2 $ & 0 &$-\pi/2 $ & 0  \\ 
 \hline
  $R_{16}$&$\pi/2 $ & $0$ & $+\pi/2 $ & 0 &$-\pi/2 $ & $\pi$  \\ 
 \hline
  $R_{17}$&$\pi/2 $ & $0$ & $-\pi/2 $ & 0 &$-\pi/2 $ & $\pi$  \\ 
 \hline
  $R_{18}$&$\pi/2 $ & $0$ & $-\pi/2 $ & 0 &$+\pi/2 $ & $\pi$  \\ 
 \hline
  $R_{19}$&$\pi/2 $ & $0$ & $-\pi/2 $ & 0 &$+\pi/2 $ & 0  \\ 
 \hline
  $R_{20}$&$\pi/2 $ & $0$ & $-\pi/2 $ & $\pi$ &$+\pi/2$ & 0  \\ 
 \hline
\end{tabular}
 \label{tbl:LSFrequencies}
 \caption{Flux Configurations that are stabilized as a function of $J_5/J$ in Fig.~\ref{fig:Phase_diag_j1111}.}
\end{table}

Next, we discuss the effects of the $J_5$ term which is an onsite term that couples to $\Gamma^5$. Since $\Gamma^\mu$ operators can be expressed as $J=3/2$ quadruple operators, $J_5$ term does not break time-reversal symmetry but it is akin to an anisotropy term. Previous studies\cite{Urban, Chua, Akram_arXiv2023} show that similar onsite terms can change the ground state vison configuration. In terms of Majorana fermion description, $J_5$ term couples the two flavors, $c$ and $d$ (see Eq.~\ref{eq:Hmf}). We consider uniform bond coupling constants, $J_1=J_2=J_3=J_4 = J$ and study the $J_5/J$ phase diagram. Our Monte Carlo simulations suggest that new vison configurations can be stabilized. However, these configurations all obey a simple rule that plaquettes with the same $n$ have the same flux configuration. Since dABL has $n=\{3,4,5,6,7,8\}$ plaquettes, there are $2^6$ vison configuration possibilities. As each configuration has a time-reversal partner and there are a total of 32 distinct possible vison configurations that follow this rule. Since the energy differences between the vison configurations can be quite small, obtaining the true ground state from Monte Carlo can be challenging due to phase separation. Therefore, we construct 32 variational configurations and compare their energies. In Fig.~\ref{fig:Phase_diag_j1111}, we show that 21 of these states can be realized as a function of $J_5/J_1$. The vison configurations of these phases are shown in Table I. Note that their time-reversal partners are also ground states. For $J_5/J_1>3.5$ all vison configurations become degenerate. This phase is unstable to confinement and has been observed in similar models with onsite anisotropy\cite{Urban}. For $J_1=J_2=J_3=J_4$ and $J_5=0$, the ground state has $\nu=\pm 2$. We find that $J_5$ term hybridizes the edge states but does not change $\nu$. Remarkably, all 21 states in Table I have the same Chern number $\nu=\pm 2$ even though they have different vison configurations.

\section{Conclusion}


We developed an exactly solvable spin liquid model on a quasicrystal lattice. The structural properties required for Kitaev type integrability are uniform coordination number $z$ throughout the lattice and the bond lattice being $z$-partite. The dual lattice (center model) of the commonly used ABL is uniformly coordinated with 4 bonds at each site. We showed that a coloring rule applied to the four dimensional simple cubic lattice generates the dABL with the required 4-partite property.
With these structural properties we used the $\Gamma$ matrix generalization of the Kitaev interactions to define our Hamiltonian. A Majorana fermion representation of the interactions reduce the Hamiltonian to a tight binding model on the quasicrystal that is coupled to a static $Z_2$ gauge field. We used both finite size quasicrystals with open boundary conditions and periodic approximants for our numerical calculations.  
\begin{figure}[!hbt]
    \centering
    \includegraphics[clip,width=0.43\textwidth]{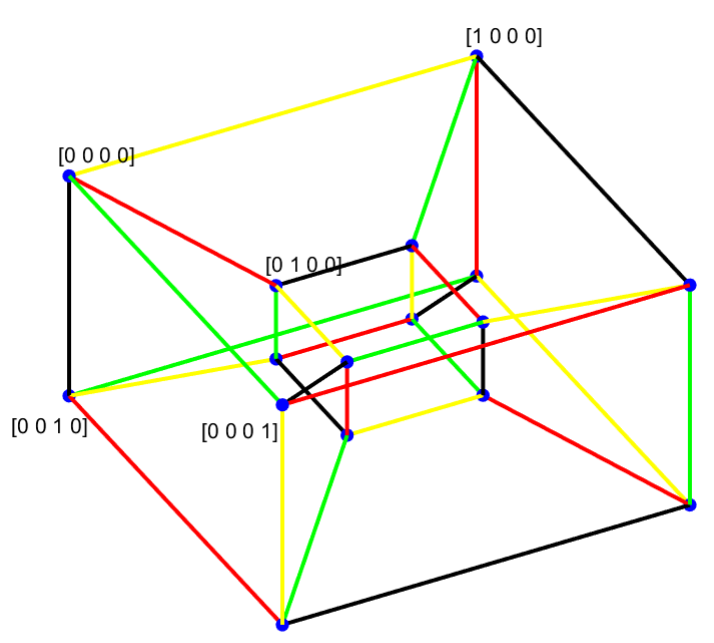}
    \caption{A  Tesseract colored with four different colors so that no square mesh has colors of the same side.}
    \label{fig:tesseract}
\end{figure}

We obtained the ground state phase diagram via the Monte Carlo simulations and variational analysis and showed that it follows a flux configuration that is an extension of the Lieb's theorem, {\it albeit} the Lieb's theorem does not directly apply here. As a function of coupling strengths, we find that the ground state can be gapped with $\nu=\pm 2$ or $\nu=0$. These phases are separated by a gapless phase. Subsequently, including an onsite term that preserves the integribility of the model, we found that 21 different vison configurations can be stabilized for isotropic exchange constants. Notably, all of these phases also share the same Chern number, $\nu=2$. Interesting future directions include extending our formalism to the emerging field of moir\'e quasicrystals.

\section{Acknowledgements}
We thank Yuan-Ming Lu and Johannes Knolle for fruitful discussions. OE acknowledges support from NSF Award 
No. DMR 2234352. MAK and MOO are supported by the TUBITAK 1001 program Grant no 122F346.

\appendix
\section{Coloring Rules }
\label{sec:Coloring}
The cut-and-project construction defines the relation between the four-dimensional simple cubic lattice and the ABL. Both the square and rhombus meshes of the ABL correspond to squares in the two-planes of the four-dimensional lattice. Therefore, we derive the coloring rules by considering a four-dimensional hypercube shown in Fig.~\ref{fig:tesseract}. We first color the intersecting edges at any corner of the tesseract with four different colors. This coloring scheme is then extended to the full simple cubic lattice by repeating reflected copies of the tesseract.

A link parallel to $\hat{e}_{n}$ connects two sites that have the same three indices $k_{m\neq n}$. We obtain the coloring rules in Table \ref{tbl:ColoringRules}.
\begin{table}
\caption{Coloring rules for the ABL so that each mesh has four edges with different colors.}
    \begin{tabular}{c|c c c}
    \multicolumn{4}{c}{$\hat{e}_0$ Bonds} \\
    \hline
     Color  & $k_1$ & $k_2$ & $k_3$  \\
     \hline
        1 & even & odd & odd \\
          & odd & even & even \\
        \hline
        2 & odd & odd & odd \\
          & even & even & even \\
          \hline
        3 & even & odd & even \\
          & odd & even & odd \\
          \hline
        4 & odd & odd & even \\
          & even & even & odd \\
    \end{tabular}
    \begin{tabular}{c|c c c}
    \multicolumn{4}{c}{$\hat{e}_1$ Bonds} \\
    \hline
     Color  & $k_0$ & $k_2$ & $k_3$  \\
     \hline
        1 & even & even & odd \\
          & odd & odd & even \\
        \hline
        2 & odd & even & odd \\
          & even & odd & even \\
          \hline
        3 & odd & even & even \\
          & even & odd & odd \\
          \hline
        4 & even & even & even \\
          & odd & odd & odd \\
    \end{tabular}
    \begin{tabular}{c|c c c}
    \multicolumn{4}{c}{$\hat{e}_2$ Bonds} \\
    \hline
     Color  & $k_0$ & $k_1$ & $k_3$  \\
     \hline
        1 & odd & odd & odd \\
          & even & even & even \\
        \hline
        2 & even & even & odd \\
          & odd & odd & even \\
          \hline
        3 & odd & even & odd \\
          & even & odd & even \\
          \hline
        4 & even & odd & odd \\
          & odd & even & even \\
    \end{tabular} 
    \begin{tabular}{c|c c c}
    \multicolumn{4}{c}{$\hat{e}_3$ Bonds} \\
    \hline
     Color  & $k_0$ & $k_1$ & $k_2$  \\
     \hline
        1 & even & odd & odd \\
          & odd & even & even \\
        \hline
        2 & odd & even & odd \\
          & even & odd & even \\
          \hline
        3 & odd & odd & odd \\
          & even & even & even \\
          \hline
        4 & odd & odd & even \\
          & even & even & odd \\
    \end{tabular}
\label{tbl:ColoringRules}
\end{table}
In all our figures, 1 $\equiv$ black, 2 $\equiv$ yellow, 3 $\equiv$ green, 4 $\equiv$ red. The color of a link in the dual lattice is the color of the link that intersects it in the ABL.

\bibliography{KitaevModel_References}

\end{document}